\documentstyle[12pt]{article}
\begin{document}
\bibliographystyle{stand}

\title{
{\rm
\rightline{hep-ph/9710433}
\ \\
\ \\}
{\Large\bf The Aharonov-Casher Effect \\
for Particles of Arbitrary Spin}
\footnote{Presented by Ya.I.Azimov at the {\it 8th International
Conference on Symmetry Methods in Physics}, Dubna, Russia,
July 28 -- August 2, 1997. Submitted to the Proceedings.}
}

\author{Ya.I.Azimov$\,{}^a$, R.M.Ryndin$\,{}^b$\\
Petersburg Nuclear Physics Institute\\
Gatchina, St.Petersburg, 188350, Russia\\
{\it e-mail: ${}^a\,$azimov@pa1400.spb.edu,}\\
{\it $~~~~~~~~~{}^b\,$ryndin@thd.pnpi.spb.ru}}
\date{}
\maketitle

\begin{abstract}
The Aharonov-Casher (AC) effect for quantum motion of a neutral
magnetized particle in the electric field is believed to be
a topological effect closely related to the known Aharonov-Bohm
(AB) effect. We study how it depends on the spin of the particle
involved. Duality of the AB and AC effects is demonstrated to exist
only for two extreme spin (and magnetic moment) projections.

Classical consideration confirms the conclusion. Motion of a
classical magnetized particle with generally oriented magnetic
moment in the AC field appears to be subjected to both forces and
torques. Only for two special orientations of the magnetic moment
(same as in the quantum case) the motion is effectively free and
similar to the AB motion of a charged particle. Thus, the AC effect
is not really topological.

Presence of higher multipoles totally destroys the possible AB-AC
duality.

\end{abstract}

\vspace{1cm}

1. The Aharonov-Bohm effect (ABE)~\cite{AB}, the first example of
purely topological effects closely related to the gauge character
of electromagnetic field, continues to be of scientific and applied
interest. Since its discovery the effect has been intensively studied
in many details, but it was only rather recently that another
topological effect in electromagnetic fields, analogous to ABE, was
suggested, namely, the Aharonov-Casher effect (ACE) in purely electric
field~\cite{AC}.

The question of existence of ACE initiated a theoretical discussion
(see refs.\cite{B,Y,G}). It had not been concluded when the neutron
beam measurements~\cite{C} confirmed ACE both qualitatively and with
rather good quantitative precision. Even more precise value for the
Aharonov-Casher phase was obtained with molecular beams in the
homogeneous electric field~\cite {H1,H2}. And the theoretical
discussion was closed by the paper of Hagen~\cite{Hag} who
demonstrated exact duality of the Dirac wave functions in the fields
of Aharonov-Bohm (AB) and Aharonov-Casher (AC).

In the present talk we consider the AB-AC duality for particles
with an arbitrary spin. In quantum theory it appears to be possible
only for two extreme spin projections normal to the field
plane~\cite{AR}.  Such a result in the 2-dimensional problem seems to
present a paradox: according to the group theory the total angular
momentum should be inessential, its role should be played by the
normal projection. For example, one could expect that the particle
motion at spin 3/2 and its normal projection 1/2 should be equivalent
to the motion at the same normal projection, but with spin 1/2.
However, the AB and AC configurations are dual for the latter case and
not dual for the former one. In order to reveal physical reasons for
such special properties of the extreme spin (and magnetic moment)
projections~\cite{AR2} we apply purely classical consideration. We
also study influence of higher multipole moments, above the charge
and magnetic moment.

2. Let us begin with quantum consideration of a spin 1/2 particle. In
the AB configuration, i.e. in the field of a thin straight-\-line
solenoid along the 3rd axis ($z$-axis) and with wave function
independent of $z$, the Dirac equation takes the form
\begin{equation}
[i\gamma_0\partial_0-\gamma_k(i\nabla_k -eA_k)-m]\Psi=0,
\end{equation}
where $k$ runs only values 1 and 2. Without invoking a special
representation of the Dirac matrices, Hagen's result~\cite{Hag} can
be understood as due to existence of the integral of motion
$\frac12\gamma^3\gamma^5$, with physical meaning of the 3rd component
of the spin axial vector.  For eigen-states
\begin{equation} \gamma^3\gamma^5\Psi_s=s\Psi_s,
\qquad s=\pm1, \end{equation}
eq.(1) can be rewritten as the equation
\begin{equation}
\left(i\gamma^\kappa\partial_\kappa-m+\frac\mu2 F'^{\alpha\beta}
\sigma_{\alpha\beta}\right)\Psi_s=0
\end{equation}
for a neutral particle with the anomalous magnetic moment $\mu$
moving in the electric field $E^k=F'^{0k}$ (see refs.~\cite{Hag,AR}):
\begin{equation}
\begin{array}{ll} \mu F'^{01}=-\frac es A_2, & \mu F'^{02}=\frac es
A_1, \\ \mu F'^{03}=0, & \mu F'^{k\ell}=0. \end{array} \end{equation}
This field corresponds just to the AC configuration, if we started
from the AB field.

For the wave function of a particle with an arbitrary spin $S>\frac12$
we use a bispinor of the $2S$ rank, symmetrical in all indices (it is
generalization of the free-particle approach of Bargmann and
Wigner~\cite{BW}). The wave equation in the AB configuration has
the Dirac-like form~\cite{AR}
\begin{equation}
\left[\beta^\alpha(i\partial_\alpha-eA_\alpha)-m\right]\Psi=0\,,
\end{equation}
where
\begin{equation}
\beta^\alpha=\frac1{2S} \sum^{2S}_{n=1}\gamma^\alpha_{(n)}\,
\end{equation}
and each Dirac matrix $\gamma^\alpha_{(n)}$ acts only on the $n$th
index of $\Psi$.

The $z$-component of spin
\begin{equation}
\xi_3=\sum^{2S}_{n=1}\left(\frac12 \gamma^3\gamma^5\right)_{(n)}
\end{equation}
is again the integral of motion for the AB configuration, with
$2S+1$ possible values from $-S$ to $+S$. The eigen-\-states $\Psi_s$
with
\begin{equation}
\xi_3\Psi_s=sS\Psi_s, \qquad s=\pm1
\end{equation}
should, evidently, be also eigen-states for any
$(\gamma^3\gamma^5)_{(n)}$:
\begin{equation}
(\gamma^3\gamma^5)_{(n)}\Psi_s=s\Psi_s.
\end{equation}
with the same $s$. Then, in the same manner as for the spin 1/2
particle, we are able to relate
eq.(5) for the AB configuration with the equation
\begin{equation}
\left( i\beta^\kappa \partial_\kappa-m-\frac\mu2 F'^{\alpha\beta}
\frac1S S_{\alpha\beta}\right)\Psi_s=0
\end{equation}
for the AC  configuration, again in the field (4). Here
\begin{equation}
S_{\kappa\delta}=iS^2[\beta_\kappa,\beta_\delta]=\sum^{2S}_{n=1}
\left(\frac12\sigma_{\kappa\delta}\right)_{(n)}
\end{equation}
are the Lorentz generators for transforming $\Psi$. Eq.(10), similar
to eq.(3), describes a neutral particle having the (anomalous)
magnetic moment $\mu$, but with the spin $S$ instead of 1/2.

Eigen-states of $\xi_3$ with eigen-values $S_3$ at
$$ |S_3|<S $$
are not eigen-states of a particular $(\gamma^3\gamma^5)_{(n)}$, and
for them the AB and AC configurations cannot be related.

Thus, we see a special role of the two spin projections, $S_3=\pm S$.
Physically, they correspond to the spin (and magnetic moment) oriented
exactly along a normal to the plane of motion (i.e. along the solenoid
or the charged thread). To understand the role of such special
orientations, deduced here in a rather formal way, we compare this
quantum result to the classical motion of a magnetized
particle~\cite{AR2}.

3. We begin with consideration of a purely classical particle
interacting with electric and magnetic fields. Comparison of
papers~\cite{B,Y,G} to each other and to earlier studies~\cite{P,Fr}
demonstrates absence of the generally accepted relativistic Lagrangian
$L_{int}$ for interaction of point-like multipoles with external
fields. To construct it we start with two requirements:

\begin{itemize} \item
In the rest-frame of the particle (at ${\bf v}=0$) $L_{ int}$ equals to
the corresponding potential energy with sign reversed; \item $L_{int}$
transforms as $dS/dt$ with $S$ being the invariant action, so it is an
invariant multiplied by \mbox{$(1-v^2)^{1/2}$}.  \end{itemize}
In this way we obtain, e.g., the following interaction Lagrangians for
the point charge and the point-like electric and magnetic dipoles (such
procedure gives, of course, the standard form to $L^{(ch)}$):
\begin{equation} L^{(ch)}({\bf x},t)=-q\phi({\bf x},t)+
q{\bf v}\cdot{\bf A(x},t)\ , \;\; \;\; \end{equation}
\begin{equation} L^{(ed)}({\bf x},t) = {\bf d}\cdot{\bf E(x},t)+{\bf
d}\cdot [{\bf vB(x},t)]\ , \; \end{equation}
\begin{equation} L^{(md)}({\bf x},t)={\bf m}\cdot {\bf B(x},t)-{\bf
m}\cdot [{\bf vE(x},t)]\ .  \end{equation}
There exists a tradition (see, e.g., refs.\cite{P,Fr}) to describe
electric and magnetic dipoles by a combined antisymmetric tensor
(similar to the field tensor {\bf(E,B)}).  However, they play very
different roles for elementary particles (e.g., ${\bf d}\neq0$ leads
to $P$ and $T$ violation). So we use two independent tensors, each one
having a special form in the particle rest-frame: $(-{\bf d}_0,0),
(0,{\bf m}_0)$. In an arbitrary system these tensors are $(-{\bf d}
(1-v^2)^{-1/2},[{\bf dv}](1-v^2)^{-1/2})$ and $([{\bf mv}](1-v^2)^{-1/2},
{\bf m}(1- v^2)^{-1/2})$ if the particle moves with the velocity {\bf v}.

Correct relativistic transformation of $L_{int}$ is provided by $q=\mbox
{inv}$, while {\bf d} and {\bf m} should change. Transition from the
rest-frame to a moving one transforms them just in the same way as the
radius vector between two space points: longitudinal components undergo
the Lorentz contraction, transversal ones do not change. Nevertheless,
we do not vary them with {\bf v}: since dipole moments of a classical
particle may change with time (in both value and direction) even in
its rest-frame, we assume them to depend directly on time but not on
velocity which is time-dependent itself.

The above interaction Lagrangians explicitly depend on velocity, so the
particle canonical momentum contains both the familiar contribution,
proportional to the vector potential, and other contributions,
proportional to other multipoles. $L^{(ch)},\ L^{(ed)}$ and $L^{(md)}$
produce additional terms
\begin{equation} \delta{\bf p}=q{\bf
A(x},t)-[{\bf dB(x},t)]+[ {\bf mE(x},t)]\ .  \end{equation}
The last two terms are sometimes considered as "hidden" momenta, but
these extra terms are really induced by the explicit dependence of
interaction on velocity.

Every multipole generates the corresponding forces of the particle
interaction with fields. For both electric and magnetic dipoles
{\bf d} and {\bf m} the forces may be presented as ${\bf f}={\bf
f}_{(0)}+\delta{\bf f}$, where respectively
\begin{equation} {\bf f}^{(ed)}_{(0)}=d_i\partial_i{\bf E}+ d_i
[{\bf v}\partial_i{\bf B}] + [\dot{\bf d}{\bf B }]\ ,\;\;
\end{equation}
\begin{equation} {\bf f}^{(md)}_{(0)}= m_i\partial_i{\bf B}- m_i
[{\bf v}\partial_i{\bf E}] - [\dot{\bf m}{\bf E}]\ .
\end{equation}
Additional terms for the two cases are
$$\delta{\bf f}^{(ed)}=[{\bf d}(\mbox{rot{\bf E}}+\dot{\bf B})]
+[{\bf dv}]\,\mbox{div{\bf B}},\;$$
$$\delta{\bf f}^{(md)}=[{\bf m}(\mbox{rot{\bf B}}-\dot{\bf E})] -
[{\bf mv}]\,\mbox{div{\bf E}}.$$
Because of the Maxwell equations, the extra term $\delta{\bf f}^{(ed)}$
vanishes, while $\delta{\bf f}^{(md)}$ appears to be expressed through
charge and current densities produced by the particle itself. So (if
not vanishing)  $\delta{\bf f}^{(md)}$ corresponds to the particle
self-interaction which influence should be included into the particle
mass and/or other parameters, but should not be explicitly present in
its equation of motion under external forces.  Note that the forces
${\bf f}_{(0)}$ may be obtained by describing dipoles {\bf d} and {\bf
m} as limiting systems of two separated opposite charges, electric $\pm
q_e$ or magnetic $\pm q_m$ (if they existed), each one subjected to the
electric or magnetic Lorentz force
\begin{equation}
{\bf f}^{(e)}= q_e{\bf E} + q_e[{\bf vB}]\ ,\;\;\; {\bf f}^{(m)}=
q_m{\bf B} - q_m[{\bf vE}]\ .  \end{equation}
${\bf f}^{(e)}$ has the familiar form and arises directly from
$L^{(ch)}$; the both forces may be obtained by the Lorentz
transformation of the force acting on a motionless charge, electric
or magnetic respectively. Strange enough, none of
papers~\cite{AC,B,Y,G} on ACE gives the total classical force (17),
though each of them correctly presents its separate terms.

There is one more point which has not been discussed at all in
connection with ACE. A particle bearing dipole moments (or higher
multipoles), even being point-like, should be considered as anisotropic.
Its orientation produces new degrees of freedom. The particle can have
non-vanishing angular momentum, it can be subjected to torques.
Expressions for the torques may be obtained from $L_{int}$ by varying
the particle orientation. For dipole moments the torques are
\begin{equation} {\bf
M}^{(ed)}=[{\bf dE}]+[{\bf d[vB}]]\ ,\; \end{equation} \begin{equation}
{\bf M}^{(md)}= [{\bf mB}]-[{\bf m[vE}]]\ . \end{equation}
In what follows we shall see  that torques are very important for
understanding properties of the particle motion.

In a similar way one can also find Lagrangians, forces and torques for
particles with any higher multipoles. We do not give them here
explicitly, but will have them in mind to consider the contribution of
higher multipoles.

Let us briefly discuss how the motion influences interaction of
multipoles with external fields. For $L_{int}$ and {\bf M} it
produces only a formal change from the real fields {\bf E} and {\bf B}
to effective ones ${\bf E}'={\bf E}+[{\bf vB}],\,\, {\bf B}'={\bf
B}-[{\bf vE}]$.  But this change is not the only physical effect, as
demonstrated by additional contributions to the canonical momentum
and force.

4. Let us apply Lagrangians (12)-(14) to the classical motion of a
particle in the Aharonov-Bohm (AB) field, i.e. in the field of the
infinite thin solenoid along the $z$-axis. Field strengths vanish
around the solenoid; the motion there looks free, without any forces
and torques. However, {\bf p} contains an extra term $\delta{\bf p(x})$
with  $\delta p_z=0$.

The situation becomes less trivial if we consider the thin solenoid as
the limit of a solenoid with a finite radius. For definiteness we
suggest the following structure of the magnetic field strength: from
the $z$-axis up to some radius the field is homogeneous and directed
along the $z$-axis; in the transient region the field conserves its
direction and diminishes its strength down to zero; the field strength
(but not potential!) is absent everywhere outside. We also assume that
initially the particle moves in the $(x,y)$ plane orthogonal to the
solenoid.

If the particle in its motion does not go through the strength region
then no force and no torque arise; the classical motion stays free. In
the strength region, as is well known, the field produces the force
${\bf f}^{(e)}$. It acts on the particle charge and tends to curve its
trajectory without driving it out of the $(x,y)$-plane. Neither
the particle orientation changes.

Let us consider the role of magnetic moment. If the dipole is normal to
the motion plane (i.e. along the $z$-axis) then, according to eqs.(17)
and (20), it is not subjected to any torque or additional force. The
particle orientation conserves.

The situation is quite different if {\bf m} has non-vanishing projection
onto the motion plane. Because of the torque ${\bf M}^{(md)}$ the
dipole {\bf m} precesses around the field direction. Precession does
not produce any additional force in the homogeneous region. However,
in the transient region it does generate the force ${\bf f}^{(md)}$
which has a non-vanishing $z$-component and draws the motion out of
the $(x,y)$-plane.  The sign of $f^{(md)}_z$ may change in the course
of precession, so the motion oscillates around its initial plane. Thus,
the particle leaves the field region with, generally, non-zero $v_z$.
Its sign and value depend on both the field structure in the transient
region and the initial orientation of {\bf m}.  Projection of {\bf m}
onto the {\bf B}-direction does not change.

The motion becomes much more complicated if the particle, in addition
to the charge and magnetic moment, has also an electric dipole moment
(EDM) or higher multipoles. In such a case the precession axis itself
changes its direction, and the particle orientation evolves in a rather
complicated way. There are additional forces and torques acting in both
transient and homogeneous regions, at any initial particle orientation.
As a result, after going through the field  the particle not only gets
out of plane, but also changes its {\bf m}-projection onto {\bf B}.

So, we see that a classical particle in the field of a long solenoid
moves differently in the presence, in addition to the charge, of some
other multipoles (even magnetic moment). Resulting motion is very
sensitive to the field structure, and, hence, the problem of the
infinitely long and thin solenoid may essentially depend on the
limiting procedure. Only for the particle with charge and magnetic
moment, exactly along the field direction, the motion has "canonical"
properties.

5. Now we consider a classical particle in the Aharonov-Casher (AC)
field, i.e., in the field of a uniformly charged straight thin thread
along the $z$-axis. Its electric strength is perpendicular to the
thread and depends only on two coordinates $x$ and $y$. We assume the
particle to be neutral, but having a magnetic moment and, may be, some
other multipoles. Initially the particle moves in the $(x,y)$-plane.

Let us begin with {\bf m} oriented along the $z$-axis. Then eq.(20)
shows that torques are absent, and the particle orientation does not
change. Furthermore, $\dot{\bf m}=0,\,\,m_x=m_y=0,$ and, due to
eq.(17), forces are also absent. The particle motion looks to be free,
but its canonical momentum, according to eq.(15), differs from the
free value $m{\bf v}(1-v^2)^{-1/2}$. The extra term $\delta{\bf p}$
has the same structure as for a charged particle in the AB field
outside the solenoid (in particular, $\delta p_z=0$). Therefore, the
whole classical description of this case in the AC field is similar
to the case of a charged particle in the AB field.

If {\bf m} has non-vanishing projection onto the $(x,y)$-plane, then
$\delta p_z\neq 0$. This means that $\delta{\bf p}$ is definitely
different here from the corresponding extra term in the AB
case.  Such motion generates the torque which compels {\bf m} to
precess around the axis $[{\bf vE}]$, initially parallel to the
$z$-direction.  Therefore, non-vanishing $\dot{\bf m}$ arises,
initially in the $(x,y)$-plane. Now, eq.(17) shows that non-zero
$\dot{\bf m}$ and projections of {\bf m} onto the $(x, y)$-plane
generate a force (with non-zero $z$-component) which changes the
{\bf v}-direction and even draws the motion out of the plane.

If one compares such classical motion in the AC configuration to the
classical motion of a charged particle in the AB field with the same
initial orientation of {\bf m}, then an essential difference can be
seen.  The AB motion is also subjected to forces and torques violating
its plane character, but only when propagating through the confined
region of the solenoid itself. In the AC field the forces and torques
act over the whole space. Thus, the AC motion with oblique orientation
of {\bf m} can never be free and never is similar to the AB motion. The
magnetic moment {\bf m} changes its orientation in a rather complicated
way: it precesses around the axis $[{\bf vE}]$ which itself changes its
direction.

Nearly of the same character is the motion of a particle carrying
also EDM or higher multipoles. But the time evolution of {\bf m} and
{\bf v} becomes more complicated. One of the reasons is that addition
of any other multipoles to the magnetic moment induces forces and
torques which affect the motion in the AC field even for the initial
{\bf m}-orientation along the $z$-axis. Therefore, possibility of
duality between the classical AB and AC motions appears totally
destroyed.

6. In conclusion, we briefly discuss correspondence between classical
and quantum descriptions of multipole interactions for a moving
particle. We have shown in the preceding Sections that similarity
between the classical AB and AC motions is possible only if the
magnetic moment was initially oriented exactly along the $z$-axis.
Such a conclusion is just the same as the result of the quantum
consideration~\cite{AR} and reveals its physical reason.

Furthermore, the classical study has shown that addition of EDM or
higher multipoles totally prevents possible AB--AC duality. Leaving
apart detailed quantum analysis of such a case we consider here only
briefly how the multipoles influence the particle wave function.

In the AB field the particle carrying only charge and magnetic moment
conserves its spin $z$-projection~\cite{Hag,AR}. This conservation
property for a Dirac particle in the AB field is expressed by
commutation of $\gamma_3 \gamma_5$ with the Dirac operator, if there
is only magnetic field directed along the $z$-axis and the wave
function depends only on $x$ and $y$. Presence of EDM induces
interaction proportional to $i\sigma_{\mu\nu}\gamma_5 F^{\mu\nu}$ ($P$
and $T$ violation does not matter here). If the field consists only of
$B_z$, we have the operator contribution $i\sigma_{12}\gamma_5$ which
does not commute with $\gamma_3\gamma_5$, and the spin orientation
cannot conserve. EDM has similar effect in the AC configuration, since
it induces the terms $i\sigma_{0k}\gamma_5$ with $k=1,2$ in the Dirac
operator.  However, the quantitative influence of EDM on the AB and AC
wave functions is different, both because of different properties of
fields {\bf B} and {\bf E} and due to their different space
distributions.  So, the AB--AC duality is excluded by the presence of
EDM.

Higher multipoles require a higher spin and complicate relativistic
consideration (see, e.g.,~\cite{AR}). But for the qualitative
understanding we may stick to the quadrupole moment in nonrelativistic
approximation.  Operator of the quadrupole moment, both electric and
magnetic, is proportional to $S_j S_k + S_k S_j -\frac {2}{3} {\bf
S}^2\delta_{jk},$ where {\bf S} is the spin operator. It is easy to
check that in both configurations, AB and AC, the interaction contains
terms not commuting with $S_3$ and not conserving the spin
$z$-component. The AB--AC duality is again impossible. So, the roles
of various multipoles in the quantum and classical descriptions agree
with each other, at least qualitatively.

One of interesting classical results is the inevitable non-plane
character of motion if the magnetic moment was initially deflected from
the $z$-axis (or in the presence of EDM or higher multipoles).
Meanwhile, the quantum wave function may still depend on
$x$ and $y$ only which is usually thought to indicate plane motion.
However, really it is not so. At the space infinity, where {\bf E}
vanishes, the canonical momentum coincides with $m{\bf
v}(1-v^2)^{-1/2}$. Let us assume that $v_z=0$ there and so {\bf p} lies
in the $(x,y)$-plane.  In the AC motion $p_z$ conserves and {\bf p}
remains in the plane. That is why the wave function depends only on two
variables. But, if {\bf m} was deflected from the $z$-axis, {\bf v}
deviates from {\bf p} (see eq.(15)) and aquires non-zero $z$-component.
Only at $|m_z|=|{\bf m}|$ the problem degenerates into a really plane
one. Detailed comparison of quantum and classical considerations here
may require to define more precisely what is the velocity (and, m.b.,
coordinate as well) for a quantum relativistic particle in the external
field.

As a result of classical consideration and its comparison to quantum
one we can note an interesting difference between ABE and ACE. In the
field of a thin solenoid the wave function of the charged particle,
while going around the solenoid, produces the known geometrical phase
independently of the form of the solenoid and orientation of the
particle magnetic moment (we consider the solenoid to be straight only
for simplicity; particle orientation is not important since the field
strength vanishes around the solenoid). On the contrary, particle
orientation is very essential in the field of a thin charged thread:
even for the straight-line thread the geometrical phase arises only
for two extreme orientations of the magnetic moment. If the thread is
curved, the magnetic moment projection onto the thread direction
cannot be fixed at all, and going around the thread is not related to
the geometrical phase. Therefore, the duality of ACE and ABE is
possible only when both the solenoid and charged thread are
straight-line. It disappears if they are bent. In difference with ACE,
the topological AB effect is directly generated by the presence of the
solenoid, independently of its form. In that sense, ACE is not a real
topological effect.

This work has been supported by the grant RFBR 96-02-18630. The authors
are grateful to A.I.Vainshtein and V.G.Zelevinsky for stimulating
discussions and to L.Michel and C.R.Hagen for discussion of the
results.


\begin{thebibliography}{xampl}
\bibitem{AB} Y.Aharonov, D.Bohm, Phys. Rev. (1959) {\bf 115}, 485.
\bibitem{AC} Y.Aharonov, A.Casher, Phys. Rev. Lett. (1984) {\bf 53},
319.
\bibitem{B} T.H.Boyer, Phys.Rev. (1987) {\bf A36}, 5083.
\bibitem{Y} Y.Aharonov, P.Pearle, L.Vaidman, Phys.Rev. (1988)
{\bf A37}, 4052.
\bibitem{G} A.S.Goldhaber, Phys.Rev.Lett. (1989) {\bf 62}, 482.
\bibitem{C} A.Cimmino et al., Phys.Rev.Lett. (1989) {\bf 63},
380.
\bibitem{H1} K.Sangster, E.A.Hinds, S.M.Barnett,
E.Riis, Phys.Rev.Lett. (1993) {\bf 71}, 3641.
\bibitem{H2} K.Sangster, E.A.Hinds, S.M.Barnett et al.,
Phys.Rev. (1995) {\bf A51}, 1776.
\bibitem{Hag} C.R.Hagen, Phys.Rev.Lett. (1990) {\bf 64}, 503.
\bibitem{AR} Ya.I.Azimov, R.M.Ryndin, Pis'ma
ZhETF (1995) {\bf 61}, 444 [JETP Lett. (1995) {\bf 61}, 453].
\bibitem{AR2} Ya.I.Azimov, R.M.Ryndin, e-print hep-ph/9707468.
\bibitem{BW} V.Bargmann, E.P.Wigner, Proc. Natl. Acad. Sci. U.S.
                         (1948) {\bf 34}, 211.
\bibitem{P} {W.Pauli. Theory of relativity. London, 1958.}
\bibitem{Fr} {Ya.I.Frenkel. Electrodynamics. ONTI publishers,
              Moscow, 1934 (in Russian).}


\end{thebibliography}
\end{document}